\documentclass[a4paper]{jpconf}
\usepackage{graphicx}
\usepackage{float}
\usepackage{lineno}
\begin{document}

\title{Study of freeze-out dynamics in STAR at RHIC Beam Energy Scan Program}

\author{Sabita Das (for the STAR collaboration)}

\address{Institute of physics, Bhubaneswar - 751005, INDIA}
\address{ Brookhaven National Laboratory, Upton, NY, 11973-5000, USA}

\ead{sabita@rcf.rhic.bnl.gov}
\begin{abstract}
The STAR detector at RHIC, due to its large uniform acceptance and excellent
particle identification capabilities, has measured a variety of
hadron species ($\pi^{\pm}$, $\emph{K}^{\pm}$, $\emph{p}$,
$\bar{\emph{p}}$,  $K^{0}_{S}$, $\Lambda$, $\bar{\Lambda}$,
$\Xi^{-}$, $\bar{\Xi}^{+}$) produced in Au+Au collisions at $\sqrt {s_{NN} }=
7.7$, 11.5, 27 and 39 GeV. These data are part of the Beam Energy Scan
(BES) program at RHIC and provide an opportunity to
measure the yields and transverse momentum spectra ($p_{T}$)  of the particles produced in
the collisions. The corresponding measurements allow to study the freeze-out properties
and dynamics of heavy ion collisions. A
 statistical thermal model analysis of particle production in BES
 energies in both grand canonical and strangeness
 canonical ensembles, is used to extract the chemical freeze-out
 parameters. The $p_{T}$ spectra, particle ratios, and the
 energy and centrality dependence of freeze-out
 parameters determined from the thermal fit of particle ratios are discussed.
\end{abstract}
\section{Introduction}
Relativistic heavy-ion collisions provide the opportunity to study
strongly interacting nuclear matter at different thermodynamic
conditions.~Quantum Chromodynamics (QCD), a fundamental theory to describe the
interactions of quarks and gluons, has anticipated the transition from
hadronic matter to 
a new state of matter called Quark-Gluon Plasma (QGP)~[1] phase at high temperature and high energy density
($\approx 1$ GeV/fm$^{3}$). The QCD phase diagram is characterized by the temperature (T) and the baryon chemical potential ($ \mu_{B}$) or the (net) baryon density ($n_{B}$),
and it contains the information about the phase boundary that
separates the QGP and hadronic phases~[2,3]. Finite temperature lattice
QCD calculations~[4] predict a cross-over from
hadronic to QGP phase at vanishing baryon chemical potential and large
T while several QCD-based calculations~[5] show that at
lower T and $\mu_{B}$ a first-order phase
transition may take place. The point in the QCD phase diagram, where
the first order phase transition ends would be the QCD critical
point~[6].  The BES program at
RHIC has been carried out with the specific aim to explore several
features of QCD phase diagram such as to search for the phase
boundaries and the location of QCD critical point by colliding nuclei
at several center-of-mass energies.
When heavy ions collide at ultra-relativistic energies, the state of
the system at the time of the final interaction is reflected by the
observed single particle spectra. The process of hadron decoupling from an
interacting system is called freeze-out. 
Freeze-out are of two types - kinetic and chemical freeze-out.
Chemical freeze-out occurs at a temperature (T$_{ch}$) when inelastic collisions cease and the
particle yields become fixed and thermal (kinetic) freeze-out occurs at a temperature
(T$_{kin}$) when elastic collisions cease and particle transverse
momenta spectra are fixed.\\
In this paper we discuss the identified transverse momentum spectra and particle ratios produced
in Au+Au collisions at $\sqrt{s_{NN}}$ = 27 GeV. Similar measurements
at other BES center-of-mass energies $\sqrt {s_{NN} }= 7.7$, 11.5, and 39 GeV are
reported in~[7]. Statistical thermal models, such as
THERMUS~[8] have been proven to be successful in describing the particle
production in heavy-ion collisions~[9, 10, 11]. Experimental particle ratios are used
in a statistical thermal model in both grand
canonical ensemble (GCE) and strangeness
canonical ensemble (SCE) approach to extract various chemical
freeze-out parameters such as chemical freeze-out temperature (T$_{ch}$),
baryon chemical potential ($ \mu_{B}$), strangeness chemical potential
($ \mu_{S}$)
and strangeness saturation factor ($\gamma_{S}$). In this study we have used mid-rapidity particle
ratios that include measured yields for charged pions ($\pi^{\pm}$),
charged kaons ($\emph{K}^{\pm}$), protons ($\emph{p}$,
$\bar{\emph{p}}$),  $K^{0}_{S}$, lambdas ($\Lambda$, $\bar{\Lambda}$)
and cascades ($\Xi^{-}$, $\bar{\Xi}^{+}$)~[7, 12]. The energy and centrality dependence of extracted chemical freeze-out
parameters in Au+Au collisions at above BES energies are studied. 
\vspace{-0.4cm} 
\section{Results}
\subsection{Identified particle spectra and ratios}
\begin{figure}[h]
\begin{minipage}{16pc}
\includegraphics[width=14pc,
height=11pc]{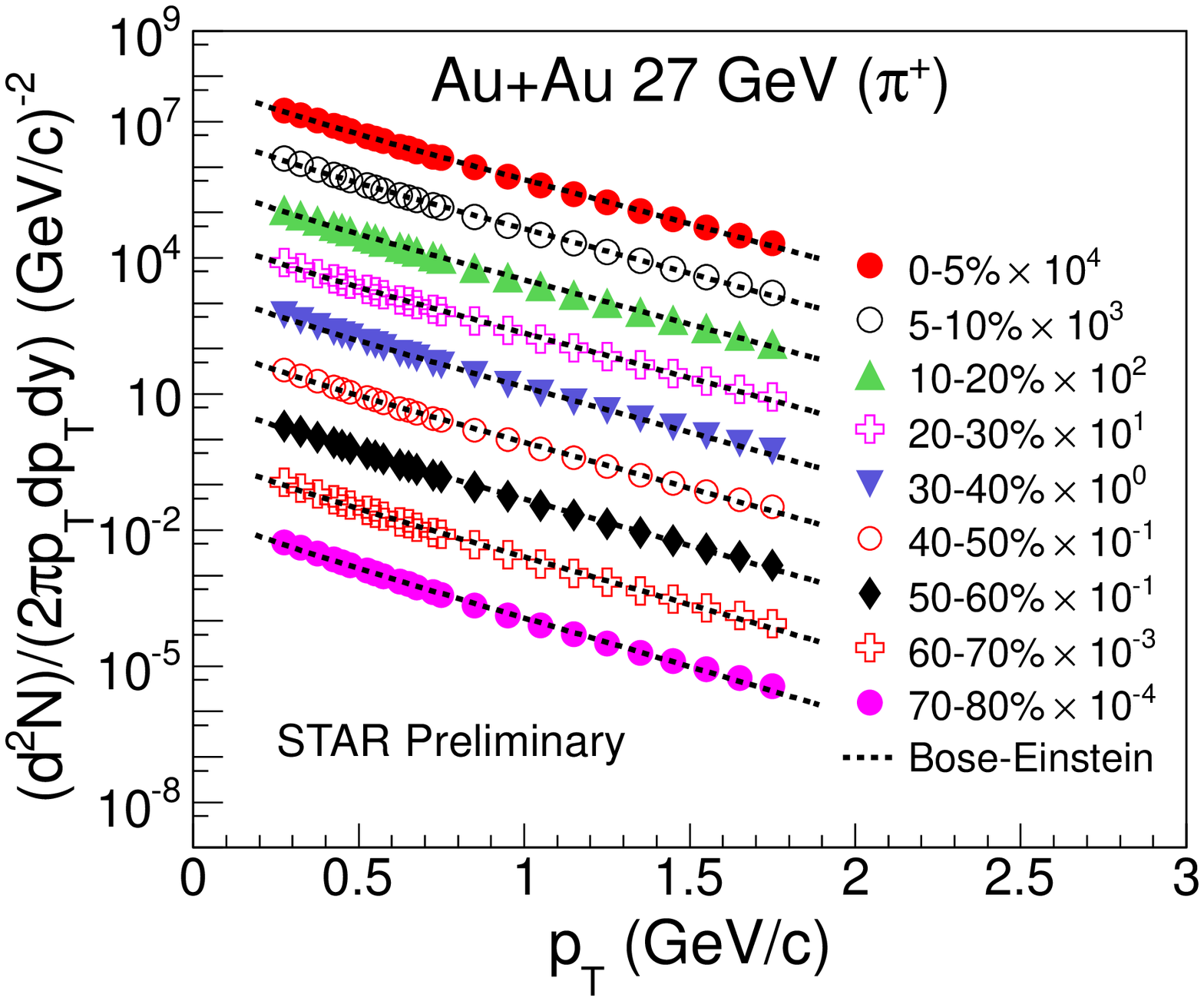}
\vspace{-0.5cm}
\caption{\label{label}Transverse momentum spectra for charged pion at
  mid-rapidity ($|y| < 0.1$) in Au+Au collisions at $\sqrt {s_{NN}
  }=27$ GeV. Errors shown here are statistical only. The lines represents the Bose-Einstein fits to the
  $p_{T}$-spectra.}
\end{minipage}\hspace{3pc}%
\begin{minipage}{16pc}
\includegraphics[width=14pc,
height=11pc]{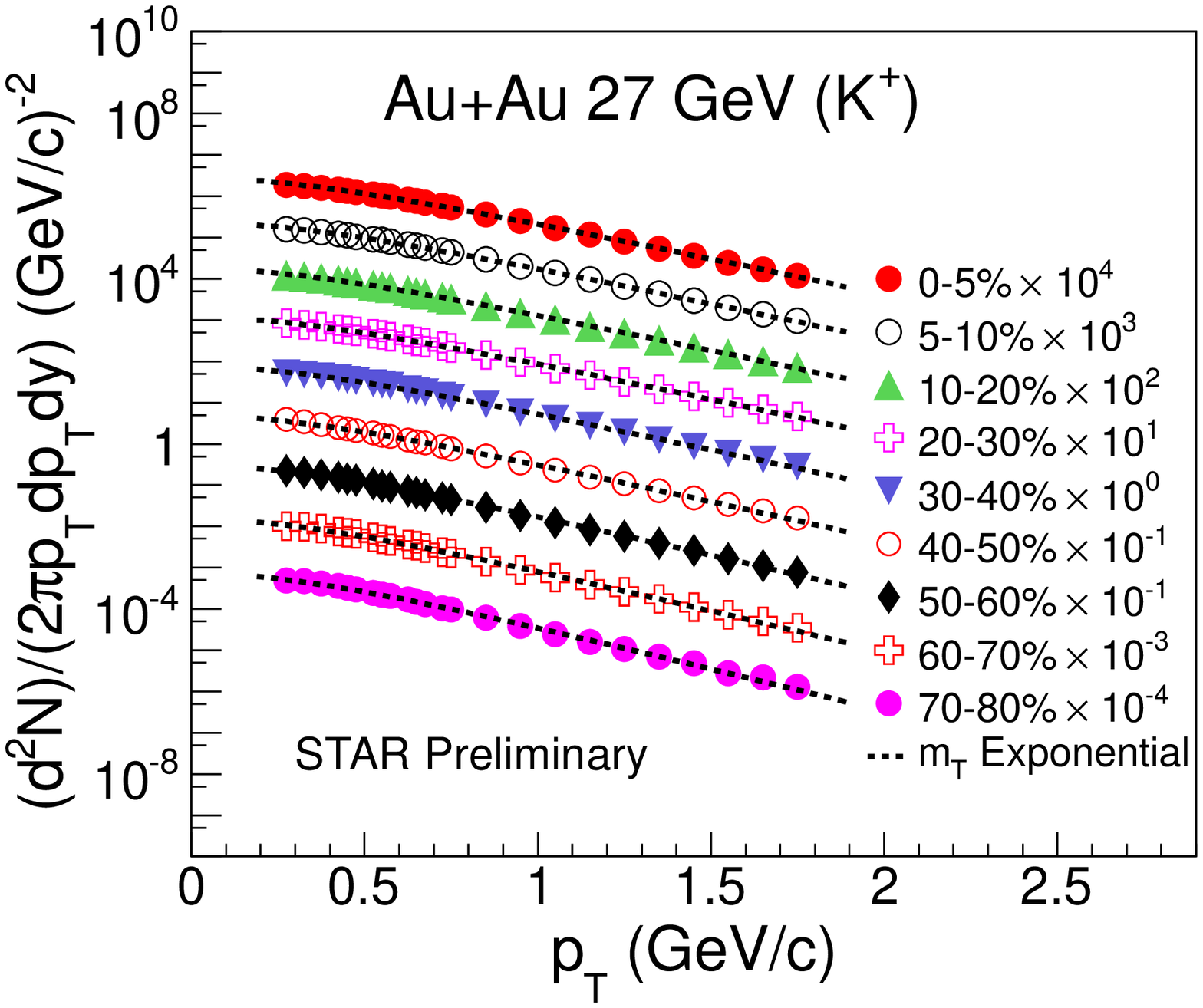}
\vspace{-0.5cm}
\caption{\label{label} Transverse momentum spectra for charged kaon
  at mid-rapidity ($|y| < 0.1$) in Au+Au collisions at $\sqrt {s_{NN}
  }=27$ GeV. Errors shown here are statistical only. The lines represents the $m_{T}$ exponential fits to the $p_{T}$-spectra.}
\end{minipage} 
\end{figure}
Figures 1 and 2 show the transverse momentum spectra of $\pi^{+}$ and
$\emph{K}^{+}$ respectively in Au+Au
collisions at $\sqrt {s_{NN} }=
27$ GeV in different centralities at mid-rapidity $|y| < 0.1$. The
$p_{T}$-integrated pion yields obtained from the Bose-Einstein fit to
$p_{T}$-spectra have been corrected for
feed-down from weak decays. The $m_{T}$-exponential fit is used to obtain
the $p_{T}$-integrated kaon yields.      
The collision-energy dependence of the
anti-particle to particle ratios in central heavy-ion collisions for $\pi^{-}/\pi^{+}$ and
$\emph{K}^{-}$/$\emph{K}^{+}$ are shown in Fig. 3. The $\pi^{-}/\pi^{+}$ ratio at lower beam
energies have values larger than unity, which could be due to
significant contributions from resonance decays (such as from $\Delta$
baryons). The $\bar{\emph{p}}$/$\emph{p}$ ratio, not shown in this
paper, increases with
increasing collision energy and approaches unity for top RHIC
energies~[21]. This indicates that at higher beam energies, the $\emph{p}$
($\bar{\emph{p}}$) production at mid-rapidity is dominated by pair
production. The $\emph{K}^{-}$/$\emph{K}^{+}$ ratio approaches unity
as collision energy increases, indicating the dominance of kaon-pair
production while at lower BES energies associated production of $\emph{K}^{+}$
dominates.
\begin{figure}
\begin{tabular}{cc}
\includegraphics[width = 2.3in,height=1.9in]{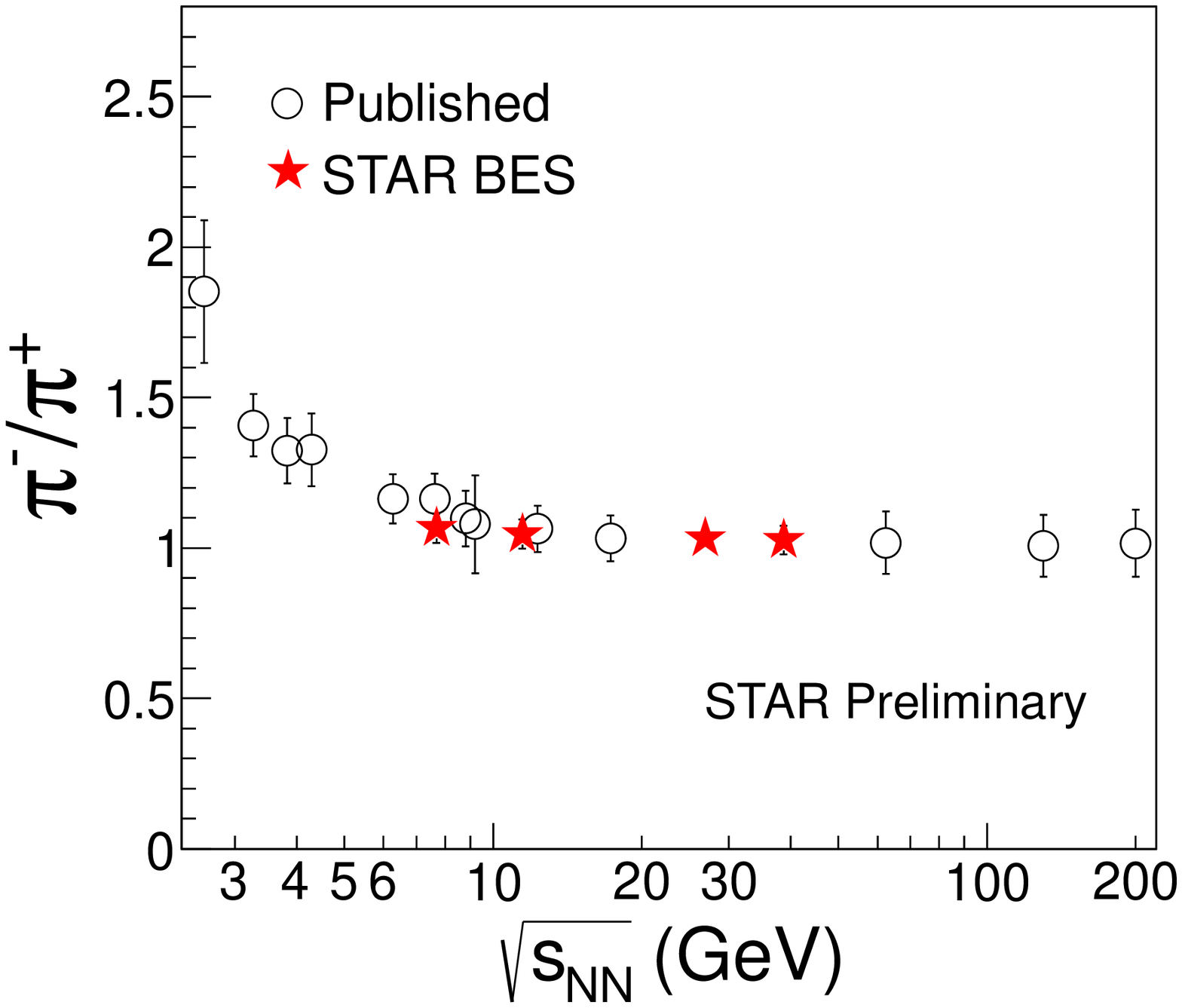} &
\hspace{2cm}
\includegraphics[width = 2.3in,height=1.9in]{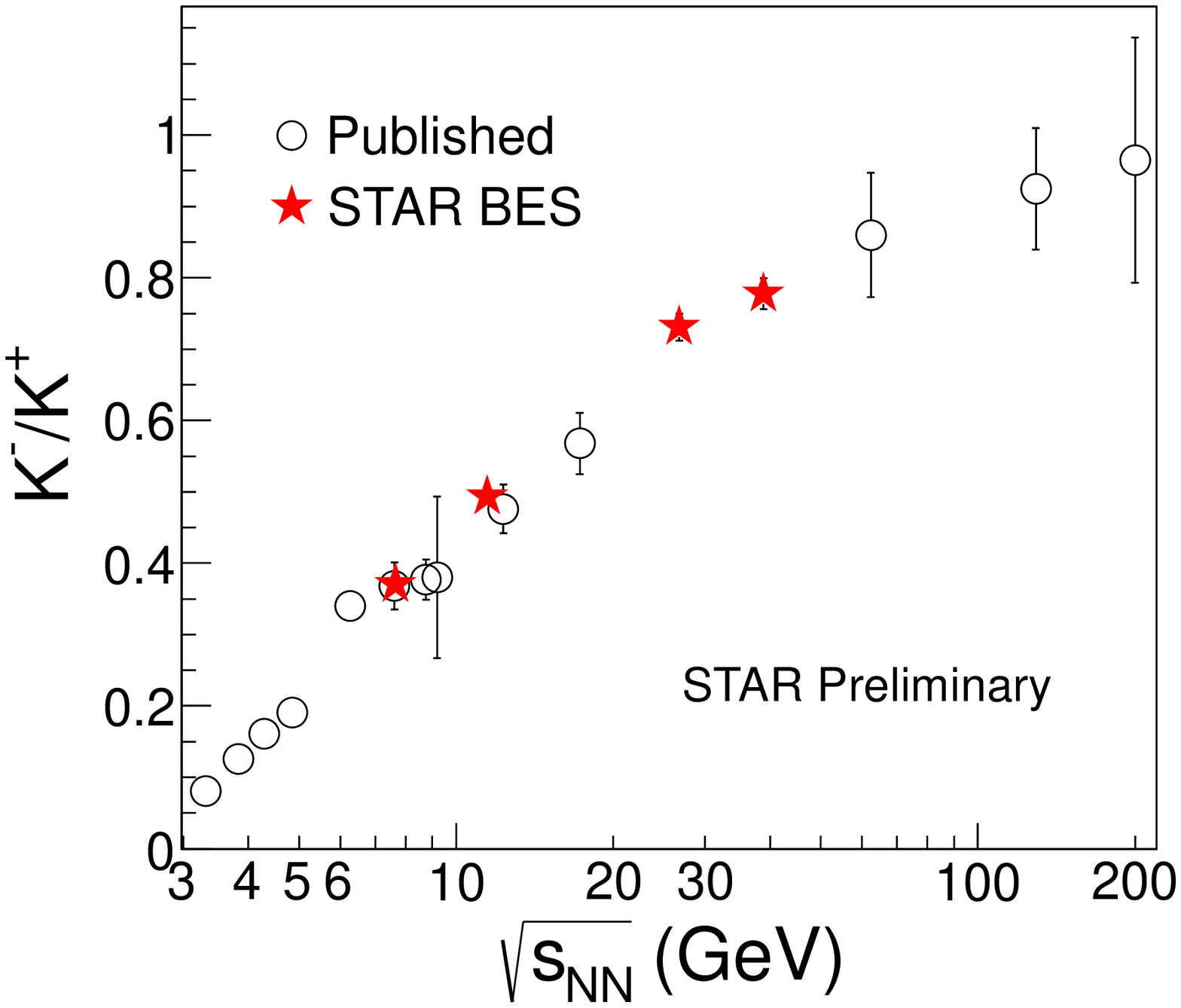}
\end{tabular}
\vspace{-0.51cm}
\caption{$\pi^{-}/\pi^{+}$ and $\emph{K}^{-}$/$\emph{K}^{+}$ ratios for 0--5\% centrality in Au+Au collisions at $\sqrt {s_{NN} }=
 27$ GeV compared with the other BES energies and previous published
 results from AGS, SPS, and RHIC~[14-22] respectively. Errors are the quadratic sum of statistical
and systematic uncertainties.}
\label{fig}
\end{figure}
\vspace{-0.02cm}
\subsection{Chemical freeze-out}
\begin{figure}
\begin{center}
\includegraphics[width=16pc, height=14pc]{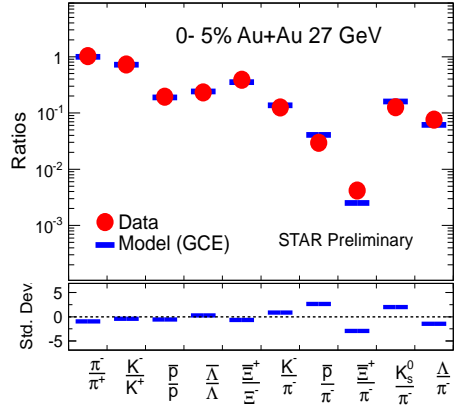}
\end{center}
\vspace{-0.6cm}
\caption{\label{label}Statistical thermal model~[8] fit to experimental 
mid-rapidity particle ratios using grand-canonical ensemble for 0--5\% centrality in Au+Au at  $\sqrt{s_{NN}} = 27$ GeV.}
\end{figure}
At chemical freeze-out, inelastic collisions among the particles stop
and hadron yields are fixed. In Au+Au collisions at $\sqrt{s_{NN} }=
27$ GeV, the measured mid-rapidity particle ratios including yields of
$\pi$, $\emph{K}$, $\emph{p}$, $K^{0}_{S}$, $\Lambda$ and $\Xi$ 
are used in THERMUS. The thermal model fit of other
BES energies at $\sqrt{s_{NN} }= 7.7$, 11.5, and 39 GeV is discussed
in~[13]. The $\pi$, $\emph{K}$,
  $\emph{p}$ yields are measured  at
  rapidity $|y| < 0.1 $ and those
  for $K^{0}_{S}$, $\Lambda$ , $\Xi$ are measured for $|y| < 0.5 $. The errors on particle ratios including yields of $\pi$,
  $\emph{K}$, $\emph{p}$, $K^{0}_{S}$, $\Lambda$, and $\Xi$, are the quadratic sum of statistical
and systematic uncertainties. Proton and anti-proton yields have not been
corrected for feed-down contributions. The $\Lambda$ yields have been corrected for the feed-down contributions from $\Xi$ and $\Xi^{0}$ weak
decays~[12]. The errors on freeze-out parameters are obtained from the
THERMUS model and errors on
particle ratios are treated as independent errors. Figure 4 shows
  the statistical thermal model fit to experimental particle ratios for 0--5\% centrality in Au+Au collisions at $\sqrt{s_{NN}} = 27$
   GeV. The data show a good agreement with the fit to model having
   $\chi^{2}/d.o.f = 23.8/6$. Figure 5(a)
   shows that the T$_{ch}$ increases
   with increasing collision energy. Figure 5(b) shows that the
   $\mu_{B}$ decreases with increasing collision energy and it
   increases when going from peripheral to central collisions at lower energies. We observe a centrality dependence of chemical freeze-out
  curve (T$_{ch}$ vs. $\mu_{B}$) at BES energies
  which was not
  observed at higher energies of Au+Au 200 GeV~[10].\\
 In contrast to GCE approach where all quantum numbers are conserved
 on average, the THERMUS model also allows for a strangeness canonical
 ensemble where only the strangeness quantum number is required to be conserved exactly where as baryon and charge quantum numbers are conserved
 on an average. It is observed that in peripheral collisions, T$_{ch}$ and $\mu_{B}$ 
 follow a different behavior in GCE and SCE at  lower BES
 energies~[13]. Further studies by using the yields for the thermal
 fit towards a more quantitative analysis for all the BES energies are
 ongoing. 
\begin{figure}
\begin{tabular}{cc}
\includegraphics[width = 2.5in,height=2.2in]{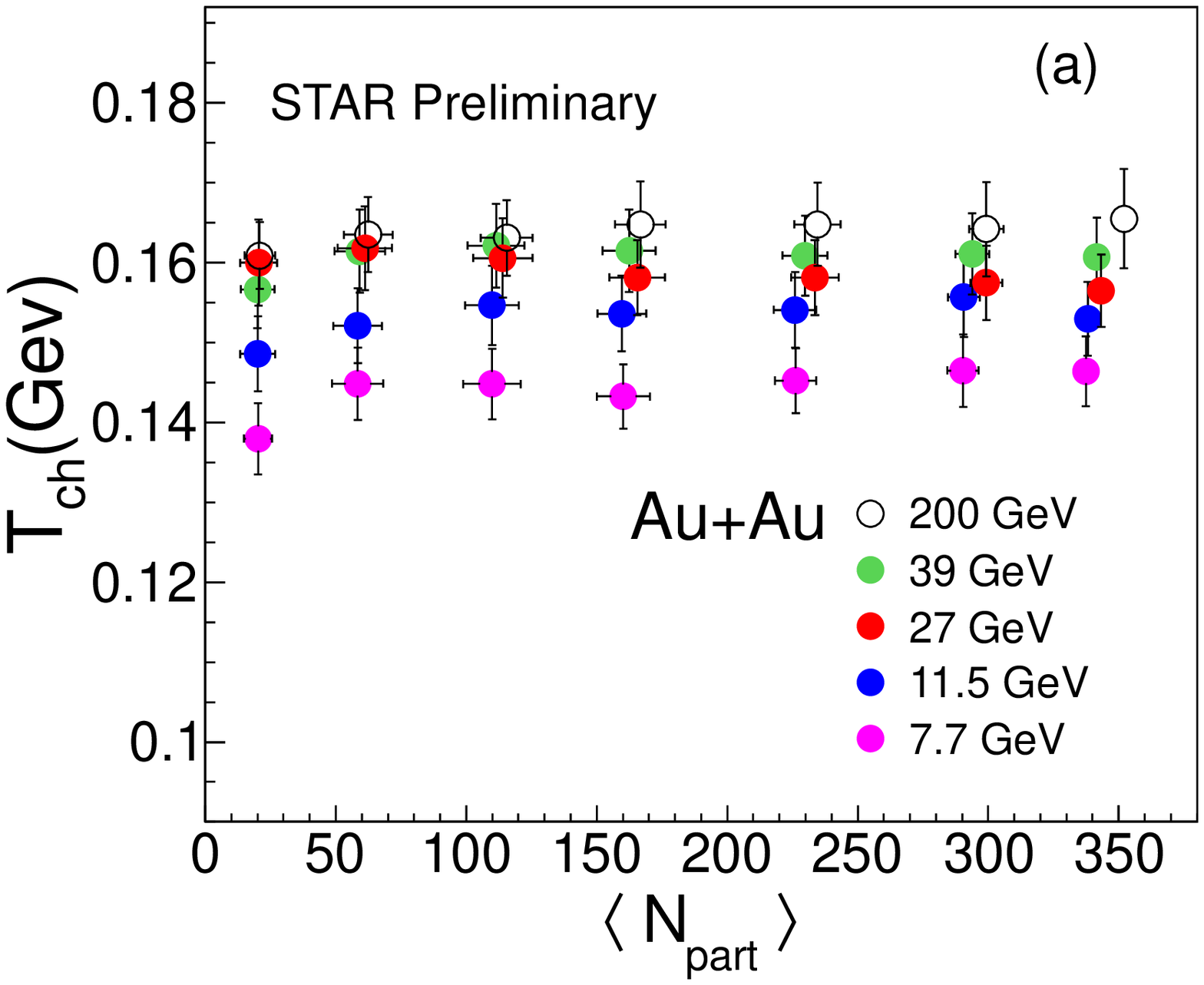} &
\hspace{2cm}
\includegraphics[width = 2.5in,height=2.2in]{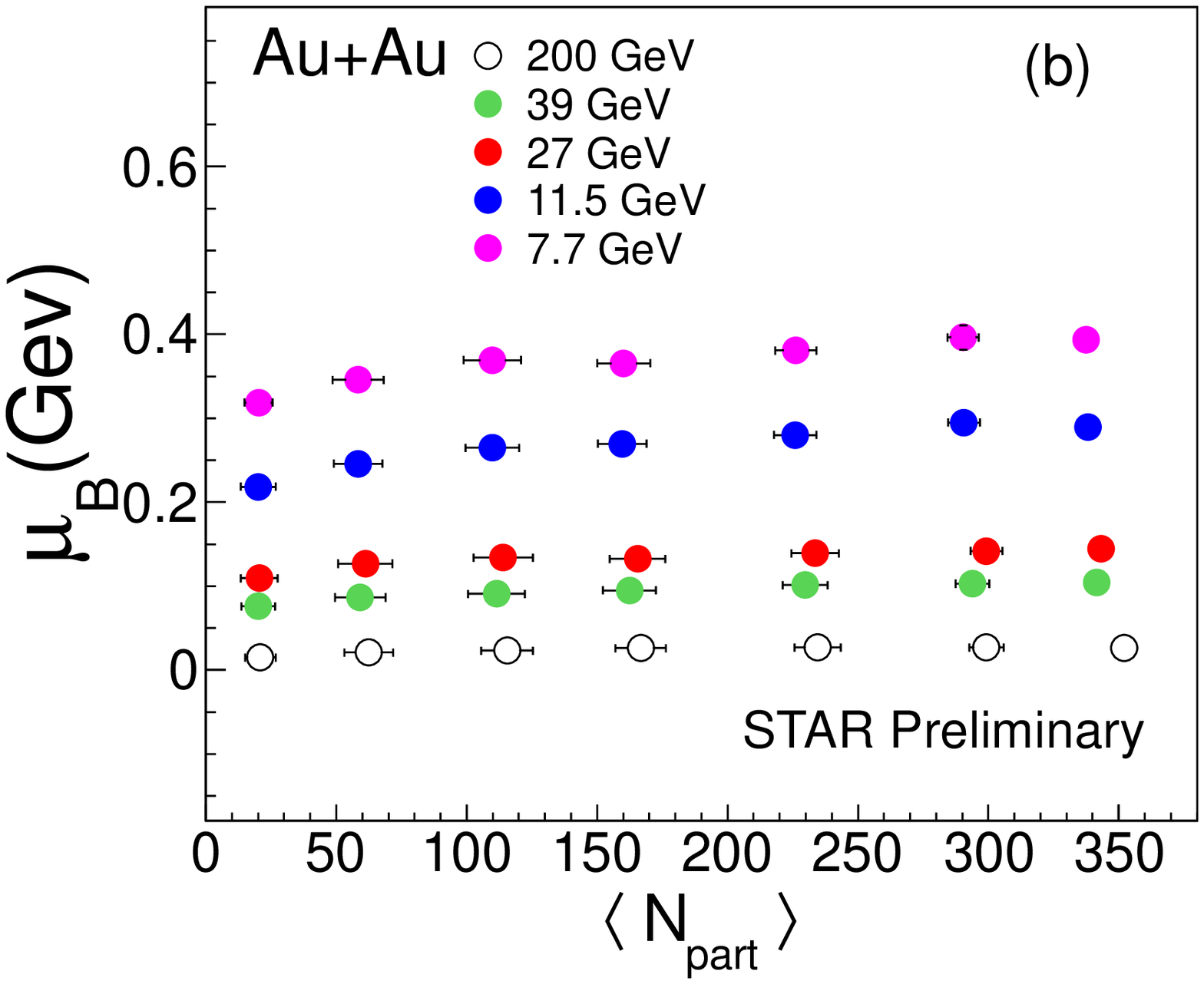}
\end{tabular}
\vspace{-0.6cm}
\caption{Results from a statistical thermal model fit~[8] for Au+Au
at $\sqrt{s_{NN}} = 7.7$, 11.5, 27 and 39 GeV. Chemical freeze-out
 temperature, baryon chemical potential, and strangeness saturation factor are
 shown as a function of the average number of participating nucleons. The 200 GeV results are taken from
Ref.~[10].}
\label{fig}
\end{figure}
\vspace{-0.4cm}
\section{Conclusions}
The study of the hadron particle ratios at BES energies in STAR within the
framework of a statistical model yields the following
conclusions. A centrality dependence of the chemical freeze-out parameters is
observed at the lower energies. Baryon chemical potential ($\mu_{B}$) range extends from 20  to 400 MeV in the QCD
phase diagram in the new BES measurements at $\sqrt{s_{NN}} = 7.7$,
11.5, 27 and 39 GeV. A detailed study using particle yields
and different choices of ensembles in the statistical model is
ongoing: this is expected to provide a better understanding of the
freeze-out dynamics.


\vspace{-0.4cm}
\section*{References}
\begin{thebibliography}{50}
 \bibitem{Julius1} Adams J \textit {et al} (STAR Collaboration) 2005 
   \textit {Nucl.Phys.} A {\bf 757} 102 
\bibitem{Julius2} Braun-Munzinger P \textit {et al} 2011 \textit
  {arXiv:1101.3167}
\bibitem{Julius3} Mohanty B 2009 \textit {Nucl. Phys.} A {\bf 830} 899C
\bibitem{Julius4} Aoki Y \textit {et al} 2006 Nature {\bf443} 675
\bibitem{Julius5} Ejiri S 2008 \textit {Phys. Rev.} D {\bf78} 074507;
  Bowman E S and Kapusta J I 2009 \textit {Phys. Rev.} C {\bf79} 015202 
\bibitem{Julius6} Gupta S \textit {et al} 2011 \textit {Science} {\bf332} 1525
\bibitem{Julius7}Kumar L (STAR Collaboration) 2011 \textit
  {J. Phys}. G \textit {Nucl. Part. Phys.} {\bf38} 124145  
\bibitem{Julius8} Cleymans J \textit {et al} 2009 \textit {Computer Physics Communications}
  {\bf180} 84
\bibitem{Julius9}Cleymans J \textit {et al} 2005 \textit {Phys.Rev.} C {\bf71} 054901
\bibitem{Julius10}Aggarwal M M \textit {et al} (STAR Collaboration) 2011
 \textit { Phys. Rev.} C {\bf83} 024901; Kumar L (STAR Collaboration) 2013
  \textit {Nucl. Phys.} A {\bf904-905} 256C
\bibitem{Julius11} Cleymans J \textit {et al} 2006 \textit {Phys. Rev.} C {\bf73} 034905
\bibitem{Julius12} Zhu X (STAR Collaboration) 2012 \textit {Acta Phys. Polon.} B
  {\bf Proc. Supp. 5} 213-218
\bibitem{Julius13}Das S (STAR Collaboration) 2013 \textit
  {Nucl. Phys.} A {\bf 904-905} 891C
\bibitem{Julius14} Ahle L\textit {et al} (E866 Collaboration and E917
  Collaboration) 2000 \textit {Phys. Lett.} B {\bf490} 53
\bibitem{Julius16}Klay J L \textit {et al} (E895 Collaboration) 2002
  \textit {Phys. Rev. Lett.} {\bf 88} 102301
\bibitem{Julius15} Ahle L \textit {et al} (E866 Collaboration and E917
  Collaboration) 2000 \textit {Phys. Lett.} B {\bf476} 1
\bibitem{Julius16} Afanasiev S V \textit {et al} (NA49 Collaboration) 2002
  \textit {Phys. Rev.} C {\bf66} 054902 
\bibitem{Julius17} Alt C \textit {et al} (NA49
Collaboration) 2008  \textit {Phys. Rev.} C {\bf77} 024903  
\bibitem{Julius18} Anticic T \textit {et al} (NA49 Collaboration) 2004
\textit {Phys.Rev.} C {\bf69} 024902
\bibitem{Julius19} Abelev B I \textit {et al} (STAR Collaboration)
  2009 \textit {Phys. Rev.} C {\bf79} 034909
\bibitem{Julius20}Adams J \textit {et al} (STAR
Collaboration) 2004 \textit {Phys. Rev. Lett.} {\bf92} 112301 
\bibitem{Julius21}Abelev B I \textit {et al} (STAR Collaboration) 2010
  \textit {Phys. Rev.} C {\bf81} 024911
\end{thebibliography}
\end{document}